\documentclass{ws-mpla}

\newcommand{\nc}{\newcommand}

\nc{\be}{\begin{equation}}

\nc{\ee}{\end{equation}}

\nc{\bea}{\begin{eqnarray}}

\nc{\eea}{\end{eqnarray}}

\nc{\xx}{\nonumber\\}

\nc{\ct}{\cite}

\nc{\la}{\label}

\nc{\eq}[1]{(\ref{#1})}

\nc{\what}[1]{\widehat{#1}}

\nc{\newcaption}[1]{\centerline{\parbox{6in}{\caption{#1}}}}

\nc{\fig}[3]{

\begin{figure}
\centerline{\epsfxsize=#1\epsfbox{#2.eps}}
\newcaption{#3. \label{#2}}
\end{figure}
}

\def\CA{{\cal A}}

\def\CH{{\cal H}}

\def\CL{{\cal L}}

\def\IB{{\hbox{{\rm I}\kern-.2em\hbox{\rm B}}}}
\def\IC{\,\,{\hbox{{\rm I}\kern-.50em\hbox{\bf C}}}}
\def\ID{{\hbox{{\rm I}\kern-.2em\hbox{\rm D}}}}
\def\IF{{\hbox{{\rm I}\kern-.2em\hbox{\rm F}}}}
\def\IH{{\hbox{{\rm I}\kern-.2em\hbox{\rm H}}}}
\def\IN{{\hbox{{\rm I}\kern-.2em\hbox{\rm N}}}}
\def\IP{{\hbox{{\rm I}\kern-.2em\hbox{\rm P}}}}
\def\IR{{\hbox{{\rm I}\kern-.2em\hbox{\rm R}}}}
\def\IZ{{\hbox{{\rm Z}\kern-.4em\hbox{\rm Z}}}}

\def\half{\frac{1}{2}}
\def\p{\partial}

\begin{document}

\markboth{Hyun Seok Yang}
{Dark Energy and Emergent Spacetime}

\catchline{}{}{}{}{}

\title{Dark Energy and Emergent Spacetime}

\author{\footnotesize Hyun Seok Yang}

\address{School of Physics, Korea Institute for Advanced Study,
Seoul 130-012, Korea \\
hsyang@kias.re.kr}



\maketitle


\begin{abstract}

A natural geometric framework of noncommutative spacetime is
symplectic geometry rather than Riemannian geometry. The Darboux
theorem in symplectic geometry then admits a novel form of the
equivalence principle such that the electromagnetism in
noncommutative spacetime can be regarded as a theory of gravity.
Remarkably the emergent gravity reveals a noble picture about the
origin of spacetime, dubbed as emergent spacetime, which is
radically different from any previous physical theory all of which
describe what happens in a given spacetime. In particular, the
emergent gravity naturally explains the dynamical origin of flat
spacetime, which is absent in Einstein gravity: A flat spacetime is
not free gratis but a result of Planck energy condensation in a
vacuum. This emergent spacetime picture, if it is correct anyway,
turns out to be essential to resolve the cosmological constant
problem, to understand the nature of dark energy and to explain why
gravity is so weak compared to other forces.

\end{abstract}

\ccode{PACS Nos.: 11.10.Nx, 11.40.Gh, 04.50.+h}

\section{Einstein Gravity and The Cosmological Constant Problem}

In general relativity, gravitation arises out of the dynamics of
spacetime being curved by the presence of stress-energy and the
equations of motion for the metric fields of spacetime are
determined by the distribution of matter and energy:
\be \la{einstein-eq}
R_{\mu\nu} - \half g_{\mu\nu} R = 8 \pi G T_{\mu\nu}.
\ee
The Einstein equations \eq{einstein-eq} describe how the geometry of
spacetime on the left-hand side (LHS) is determined dynamically, at
first sight, in harmony with matter fields on the right-hand side
(RHS).

However there is a deep conflict between the spacetime geometry
described by general relativity and matter fields described by
quantum field theory (QFT). If spacetime is flat, i.e., $g_{\mu\nu}
= \eta_{\mu\nu}$, the LHS of Eq.\eq{einstein-eq} identically
vanishes and so the energy-momentum tensor of matter fields should
vanish, i.e., $T_{\mu\nu} = 0$. In other words, a flat spacetime is
free gratis, i.e., costs no energy. But the concept of empty space
in Einstein gravity is in an acute contrast to the concept of vacuum
in QFT where the vacuum is not empty but full of quantum
fluctuations. As a result, the vacuum is extremely heavy whose
weight is roughly of Planck mass, i.e., $\rho_{\rm vac}
\sim M_P^4$.

The conflict rises to the surface that gravity and matters respond
differently to the vacuum energy and perplexingly brings about the
notorious cosmological constant (CC) problem. Indeed the clash
manifests itself as a mismatch of symmetry between gravity and
matters.\ct{tpad} To be precise, if we shift a matter Lagrangian
$\CL_M$ by a constant $\Lambda$, that is,
\be \la{shift}
\CL_M \to \CL_M - 2 \Lambda,
\ee
it results in the shift of the matter energy-momentum tensor by
$T_{\mu\nu} \to T_{\mu\nu} - \Lambda g_{\mu\nu}$ in the Einstein
equation \eq{einstein-eq} although the equations of motion for
matters are invariant under the shift \eq{shift}. Definitely the
$\Lambda$-term in Eq.\eq{shift} will appear as the CC in Einstein
gravity and it affects the spacetime structure. For example, a flat
spacetime is no longer a solution of Eq.\eq{einstein-eq}.

Let us sharpen the problem arising from the conflict between the
geometry and matters. In QFT there is no way to suppress quantum
fluctuations in a vacuum. Fortunately the vacuum energy due to the
quantum fluctuations, regardless of how large it is, does not make
any trouble to QFT thanks to the symmetry \eq{shift}. However the
general covariance requires that gravity couples universally to all
kinds of energy. Therefore the vacuum energy $\rho_{\rm vac}
\sim M_P^4$ will induce a highly curved spacetime whose curvature
scale $R$ would be $\sim M_P^2$ according to Eq.\eq{einstein-eq}. If
so, the QFT framework in the background of quantum fluctuations must
be broken down due to a large back-reaction of background spacetime.
But we know that it is not the case. The QFT is well-defined in the
presence of the vacuum energy $\rho_{\rm vac} \sim M_P^4$ and the
background spacetime still remains flat. So far there is no
experimental evidence for the vacuum energy to really couple to
gravity \footnote{Contrary to the Einstein gravity, the emergent
gravity reveals a completely different picture \ct{origin}: A vacuum
itself does not gravitate and only fluctuations around the vacuum
generate gravitational fields. After some thought, one may then
realize that the emergent gravity will tame the previous conflict
between the geometry and matters. See Sec. 3.} while it is believed
that the vacuum energy is real as experimentally verified by the
Casimir effect.

Which side of Eq.\eq{einstein-eq} is the culprit giving rise to the
incompatibility ? After consolidating all the suspicions inferred
above, we throw a doubt on the LHS of Eq.\eq{einstein-eq},
especially, on the result that a flat spacetime is free gratis,
i.e., costs no energy. It would be remarked that such a result is
not compatible with the inflation scenario either which shows that a
huge vacuum energy in a highly nonequilibrium state is required to
generate an extremely large spacetime. Note that Einstein gravity is
not completely background independent since it assumes the prior
existence of a spacetime manifold.\footnote{Here we refer to a
background independent theory where any spacetime structure is not
{\it a priori} assumed but defined from the theory.} In particular,
the flat spacetime is a geometry of special relativity rather than
general relativity and so it is assumed to be {\it a priori} given
without reference to its dynamical origin. This reasoning implies
that the negligence about the dynamical origin of flat spacetime
defining a local inertial frame in general relativity might be a
core root of the incompatibility inherent in Eq.\eq{einstein-eq}.

All in all, it is tempted to infer that a flat spacetime may be not
free gratis but a result of Planck energy condensation in a vacuum.
Surprisingly, if that inference is true, it appears as the H\'oly
Gr\'ail to cure several notorious problems in theoretical physics;
for example, to resolve the CC problem, to understand the nature of
dark energy and to explain why gravity is so weak compared to other
forces. After all, the problem is to formulate a physically viable
theory (i.e., a background independent theory) to correctly explain
the dynamical origin of flat spacetime. It turns out
\ct{origin,inst,e-gravity} that the emergent gravity from
noncommutative (NC) geometry precisely realizes the desired
property.

\section{Noncommutative Spacetime and Emergent Gravity}

Consider the electromagnetism on a D-brane whose data are given by
$(M, g, B)$ where $M$ is a smooth manifold equipped with a metric
$g$ and a symplectic structure $B$. The dynamics of $U(1)$ gauge
fields on the D-brane is described by open string field theory whose
low energy effective action is given by DBI action. The DBI action
predicts two important results: (I) The triple $(M, g, B)$ comes
only into the combination $(M, g + \kappa B)$, which embodies a
generalized geometry continuously interpolating between symplectic
geometry $(|\kappa Bg^{-1}| \gg 1)$ and Riemannian geometry
$(|\kappa Bg^{-1}| \ll 1)$. (II) The electromagnetic force $F=dA$
appears only as the deformation of symplectic structure $\Omega(x) =
(B + F)(x)$. Including the $U(1)$ gauge fields, the data of
`D-manifold' are given by $(M, g, \Omega)= (M, g + \kappa
\Omega)$.

Consider an another D-brane whose `D-manifold' is described by
different data $(N, G, B) = (N, G + \kappa B)$. A question is
whether there exist a diffeomorphism $\phi : N \to M$ such that
$\phi^* (g + \kappa \Omega) = G + \kappa B$ on $N$. In order to
answer the question, note that a D-brane whose worldvolume $M$
supports a symplectic structure $B$ respects the so-called
$\Lambda$-symmetry in addition to Diff$(M)$, defined by
\be \la{Lambda}
(B, A) \to (B - d\Lambda, A + \Lambda),
\ee
where the gauge parameter $\Lambda$ is a one-form on $M$. Consider
the symmetry transformation which is a combination of the
$\Lambda$-transformation \eq{Lambda} followed by a diffeomorphism
$\phi : N \to M$. The action transforms the ``DBI metric" $g +
\kappa B$ on $M$ according to $g + \kappa B \to \phi^* (g + \kappa
\Omega)$. The crux is that there ``always" exists a diffeomorphism
$\phi : N \to M$ such that $\phi^* (\Omega) = B$, which is precisely
the Darboux theorem in symplectic geometry. Then we arrive at a
remarkable fact that two different DBI metrics $g + \kappa \Omega$
and $G + \kappa B$ are diffeomorphic each other, i.e., $\phi^* (g +
\kappa \Omega) = G + \kappa B$, where $G = \phi^* (g)$. Note that
this property holds for any pair $(g,B)$ of Riemannian metric $g$
and symplectic structure $B$.

Since the symplectic structure $B$ is nondegenerate at any point $y
\in M$, we can invert this map to obtain the map $\theta
\equiv B^{-1}: T^* M \to TM$. This cosymplectic structure $\theta
\in \bigwedge^2 TM$ is called the Poisson structure of $M$ which
defines a Poisson bracket $\{\cdot,\cdot\}_\theta$. A NC spacetime
then arises from quantizing the symplectic manifold $(M, B)$ with
its Poisson structure, treating it as a quantum phase space, i.e.,
\begin{equation} \label{vacuum-spacetime}
\langle B_{ab} \rangle_{\rm vac} = (\theta^{-1})_{ab} \;\; \Leftrightarrow
\;\; [y^a, y^b]_\star = i \theta^{ab}.
\end{equation}

The above argument reveals a superb physics in NC spacetime. There
``always" exists a coordinate transformation to locally eliminate
the electromagnetic force $F=dA$ as long as a manifold $M$ supports
a symplectic structure $B$, i.e., $M$ becomes a NC space. That is,
the NC spacetime admits a novel form of the equivalence principle,
known as the Darboux theorem, for the geometrization of the
electromagnetism. Since it is always possible to find a coordinate
transformation $\phi \in {\rm Diff}(M)$ such that $\phi^* (B+F) =
B$, the relationship $\phi^* \big(g + \kappa (B+F) \big) = G +
\kappa B$ clearly shows that the electromagnetic fields in the DBI
metric $g + \kappa (B+F)$ now appear as a new metric $G = \phi^*
(g)$. One may also consider the inverse relationship $\phi_*(G +
\kappa B)= g + \kappa (B+F)$ which implies that a nontrivial metric
$G$ in a vacuum \eq{vacuum-spacetime} can be interpreted as an
inhomogeneous condensation of gauge fields on a `D-manifold' with
metric $g$. It might be pointed out that the relationship in the
case of $\kappa = 2 \pi \alpha' = 0$ is the familiar diffeomorphism
in Riemannian geometry and so it says nothing marvelous.

We observed that the Darboux theorem for symplectic structures
immediately leads to the diffeomorphism between two different DBI
metrics. In terms of local coordinates $\phi: y \mapsto x = x(y)$,
the diffeomorphism then reads as\ct{origin}
\be \la{dbi-iso}
(g + \kappa \Omega)_{\alpha\beta} (x) = \frac{\p y^a}{\p x^\alpha}
\Bigl(G_{ab}(y)+  \kappa B_{ab}(y) \Bigr) \frac{\p y^b}{\p x^\beta}
\ee
where $\Omega = B + F$ and
\be \la{gauge-metric}
G_{ab}(y) = \frac{\p x^\alpha}{\p y^a} \frac{\p x^\beta}{\p y^b}
g_{\alpha\beta}(x).
\ee
Eq.\eq{dbi-iso} conclusively shows how NC gauge fields manifest
themselves as a spacetime geometry. To expose the intrinsic
connection between NC gauge fields and spacetime geometry, let us
represent the coordinate transformation in Eq.\eq{dbi-iso} as
\be \la{cov-cod}
x^a(y) = y^a + \theta^{ab} \widehat{A}_b(y).
\ee
Note that $F(x) = 0$ or equivalently $\widehat{A}_a(y) = 0$
corresponds to $G_{ab} = g_{ab}$ as it should be. Clearly
Eq.\eq{gauge-metric} embodies how the metric on $M$ is deformed by
the presence of NC gauge fields.\ct{e-gravity}

Now we will elucidate how the Darboux theorem in symplectic geometry
materializes as a novel form of the equivalence principle such that
the electromagnetism in NC spacetime can be regarded as a theory of
gravity. An important observation is that, for a given Poisson
algebra $(C^\infty(M), \, \{\cdot,\cdot\}_\theta)$, there exists a
natural map $C^\infty(M) \to TM: f \mapsto X_f$ between smooth
functions in $C^\infty(M)$ and vector fields in $TM$ such that
\be \la{ham-vec}
X_f (g) = \{ g, f \}_\theta
\ee
for any $g \in C^\infty(M)$. Actually the $1-1$ correspondence
between $C^\infty(M)$ and $\Gamma(TM)$, i.e., vector fields in $TM$,
is the Lie algebra homomorphism in the sense that
\be \la{lie-homo}
X_{\{ f, g \}_\theta} = - [X_f, X_g].
\ee

The covariant and background independent coordinates $x^a(y) \in
C^\infty(M)$ in Eq.\eq{cov-cod} are smooth functions in $M$ and so
they can be mapped to vector fields in $TM$ according to
Eq.\eq{ham-vec}. Let us denote the vector fields as $V_a \in
\Gamma(TM)$ and their dual vector space as $D^a \in \Gamma(T^*M)$.
It can be shown \ct{origin} that the vector fields $V_a$ and the
one-forms $D^a$ are related to the orthonormal frames (vielbeins)
$E_a \in \Gamma(TM)$ and $E^a \in \Gamma(T^*M)$ by $V_a = \lambda
E_a$ and $E^a = \lambda D^a$, respectively, where $\lambda^2 = \det
V_a^b$. Therefore we see that the Darboux theorem in symplectic
geometry implements a deep principle to realize a Riemannian
manifold as an emergent geometry from NC gauge fields \eq{cov-cod}
through the correspondence \eq{ham-vec} whose metric is given
by\ct{large-n}
\be \la{emergent-metric}
ds^2 = g_{ab} E^a \otimes E^b = \lambda^2 g_{ab} D^a_\mu D^b_\nu
dy^\mu \otimes dy^\nu.
\ee

Let us point out that the emergent gravity can be generalized to
full NC gauge fields in two different ways which are dual to each
other. The first notable point is that the correspondence
\eq{ham-vec} between a Poisson algebra $(C^\infty(M),
\{\cdot,\cdot\}_\theta)$ and vector fields in $\Gamma(TM)$ can be
generalized to a NC C*-algebra $(\CA_\theta, [\cdot, \cdot]_\star)$
by considering an adjoint action of NC gauge fields
$\widehat{D}_a(y) \in \CA_\theta$ as
follows\ct{origin,inst,e-gravity}
\be \la{nc-vector}
ad_{\widehat{D}_a} [\widehat{f}] (y) \equiv  - i [
\widehat{D}_a(y), \widehat{f}(y) ]_\star = V_a^\mu (y) \frac{\partial f(y)}{\partial y^\mu}
+ {\cal O}(\theta^3)
\ee
where the leading term exactly recovers the vector fields in
Eq.\eq{ham-vec}. The second point is that every NC space can be
represented as a theory of operators in a Hilbert space $\CH$, which
consists of NC C*-algebra $\CA_\theta$, and any operator in
$\CA_\theta$ or any NC field can be represented as a matrix whose
size is determined by the dimension of $\CH$. For the Moyal NC space
as an example of Eq.\eq{vacuum-spacetime}, one gets $N \times N$
matrices in the $N \to \infty$ limit. In this sense, the emergent
geometry arising from the vector fields in the NC space
\eq{vacuum-spacetime} can be understood as a dual geometry of large
$N$ matrices in $\CH$ according to the large $N$ duality or AdS/CFT
correspondence.\ct{origin,large-n}

\section{Emergent Spacetime and Dark Energy}

We are ready to clarify how the emergent gravity outlined in the
previous section reveals a noble picture about the origin of
spacetime, dubbed as emergent spacetime, which is radically
different from any previous physical theory all of which describe
what happens in a given spacetime. We will consider the Moyal NC
space in Eq.\eq{vacuum-spacetime}, i.e., $M = \IR^{2n}$ with $B =
{\rm constant}$, for simplicity.\footnote{See Ref. 2 for a general
NC spacetime. Interestingly the emergent gravity based on NC
geometry endows a natural concept of ``emergent time" since a
symplectic manifold $(M, B)$ always admits a Hamiltonian dynamical
system on $M$ defined by a Hamiltonian vector field $X_H$, i.e.,
$\iota_{X_H}B = dH$, described by $\frac{df}{dt} = X_H(f) =
\{f,H \}_\theta$ for any $f \in C^\infty (M)$. In a general case where
the symplectic structure is changing along the dynamical flow as in
Eq.\eq{dbi-iso}, the dynamical evolution must be momently defined on
every local Darboux chart as expected on a general ground.} In this
case, the background NC space \eq{vacuum-spacetime},  where every
physical processes take place, is described by $x^a(y) = y^a$ in
Eq.\eq{cov-cod}. The corresponding vector fields are given by $V_a =
\frac{\partial}{\partial y^a}$ according to Eq.\eq{ham-vec} or
Eq.\eq{nc-vector} and so the metric for the background
\eq{vacuum-spacetime} is flat, i.e., $g_{ab} = \delta_{ab}$.

We will take the picture prescribed in the footnote c for emergent
time. Then one can see that the background
\eq{vacuum-spacetime} gives rise to a flat spacetime, i.e.,
$g_{\mu\nu} = \eta_{\mu\nu}$.\ct{origin} A tangible difference from
Einstein gravity arises at this point. The flat spacetime is not
coming from an empty space but emerges from a uniform condensation
of gauge fields in a vacuum \eq{vacuum-spacetime}. Note that the
uniform condensation \eq{vacuum-spacetime} should appear as the
energy-momentum in Einstein gravity and so a flat spacetime is not
allowed.

Since gravity emerges from NC gauge fields, the parameters,
$g_{YM}^2$ and $|\theta|$, defining, e.g., 4-dimensional NC gauge
theory should be related to the Newton constant $G$ in emergent
gravity. A simple dimensional analysis shows that $\frac{G
\hbar^2}{c^2} \sim g_{YM}^2 |\theta|$. This relation
immediately leads to the fact that the energy density of the vacuum
\eq{vacuum-spacetime} is  $\rho_{\rm vac} \sim |B_{ab}|^2
\sim M_P^4$ where $M_P = (8\pi G)^{-1/2} \sim 10^{18} GeV$ is the
Planck mass. Therefore the emergent gravity finally reveals a
remarkable picture that the huge Planck energy $M_P$ is actually the
origin of a flat spacetime. In other words, a flat spacetime is not
free gratis but a result of Planck energy condensation in a vacuum.
Hence a vacuum energy does not gravitate unlike as Einstein gravity.\ct{cc-problem}

If gauge field fluctuations are turned on, i.e., $F \neq 0$, the
spacetime metric will no longer be flat as can be seen from
Eq.\eq{emergent-metric}. However, the flat spacetime should be very
robust against any perturbations since the vacuum
\eq{vacuum-spacetime} was triggered by the Planck energy
condensation, the maximum energy in Nature. Then the gravity
generated by the deformations of the background
\eq{vacuum-spacetime} will be very weak since the spacetime vacuum
is very solid with a stiffness of the Planck scale. So the dynamical
origin of flat spacetime is intimately related to the weakness of
gravitational force.

Recall that the Planck energy condensation in the vacuum
\eq{vacuum-spacetime} causes the spacetime to be NC and the NC
spacetime is the essence of emergent gravity. The UV/IR mixing in
the NC spacetime then implies that any UV fluctuations of the Planck
scale $L_P$ will be necessarily paired with IR fluctuations of a
typical scale $L_H$. These vacuum fluctuations around the flat
spacetime will add a tiny energy $\delta \rho$ to the vacuum so that
the total energy density is equal to $\rho \sim M_P^4 + \delta \rho$. A
simple dimensional analysis and a symmetry consideration, e.g., the
cosmological principle, lead to the estimation of the vacuum
fluctuation as $\rho \sim M_P^4 \big(1 + \frac{L_P^2}{L_H^2}
\big)$.\ct{tpad} Since the first term in $\rho$ does not gravitate, the
second term $\delta \rho \sim \frac{1}{L_P^2 L_H^2}$ will be a
leading contribution to the deformation of spacetime curvature,
leading to possibly a de Sitter phase. Interestingly this energy of
vacuum fluctuations, $\delta \rho \sim \frac{1}{L_P^2 L_H^2}$, is in
good agreement with the observed value of current dark
energy.\ct{tpad,cc-problem}



\begin{thebibliography}{0}


\bibitem{tpad} T. Padmanabhan, Gen. Rel. Grav. {\bf 40}, 529 (2008).


\bibitem{origin} H. S. Yang, Emergent Spacetime and The Origin of Gravity, {\tt arXiv:0809.4728}.



\bibitem{inst} H. S. Yang, Instantons and Emergent Geometry, {\tt hep-th/0608013}.



\bibitem{e-gravity} H. S. Yang, Emergent Gravity from Noncommutative Spacetime, {\tt hep-th/0611174}.



\bibitem{large-n} H. S. Yang, {\tt arXiv:0704.0929}.



\bibitem{cc-problem} H. S. Yang, {\tt arXiv:0711.2797}.


\end{thebibliography}
\end{document}